\documentclass[letterpaper,12pt]{article}
\usepackage[margin=0.75in]{geometry}
\usepackage{xcolor, natbib}
\usepackage{amssymb,amsmath,amsfonts}
\usepackage{titlesec}

\numberwithin{equation}{section}

\usepackage{subfig, float, url, rotating}
\usepackage{enumitem}
\usepackage{bm, bbm} 
\usepackage{booktabs} 
\usepackage{multicol, multirow} 
\usepackage{mathrsfs} 
\usepackage{scalerel,stackengine} 

\renewcommand{\epsilon}{\varepsilon}

\stackMath

\renewcommand{\epsilon}{\varepsilon}

\renewcommand{\tilde}{\widetilde}

\allowdisplaybreaks

\begin{document}
\title{Monte Carlo sampling of flexible protein structures: an application to the SARS-CoV-2 omicron variant}






\author{Samuel W.K. Wong\footnote{Address for correspondence: samuel.wong@uwaterloo.ca} \\
Department of Statistics and Actuarial Science \\
University of Waterloo, Waterloo, ON, Canada}

\maketitle

\begin{abstract}
Proteins can exhibit dynamic structural flexibility as they carry out their functions, especially in binding regions that interact with other molecules. For the key SARS-CoV-2 spike protein that facilitates COVID-19 infection, studies have previously identified several such highly flexible regions with therapeutic importance. However, protein structures available from the Protein Data Bank are presented as static snapshots that may not adequately depict this flexibility, and furthermore these cannot keep pace with new mutations and variants. In this paper we present a sequential Monte Carlo method for broadly sampling the 3-D conformational space of protein structure, according to the Boltzmann distribution of a given energy function. Our approach is distinct from previous sampling methods that focus on finding the lowest-energy conformation for predicting a single stable structure. We exemplify our method on the SARS-CoV-2 omicron variant as an application of timely interest. Our results identify sequence positions 495--508 as a key region where omicron mutations have the most impact on the space of possible conformations, which coincides with the findings of other preliminary studies on the binding properties of the omicron variant. \\
\end{abstract}

%


\section{Introduction}

SARS-CoV-2 is the virus that causes COVID-19 and its genome was rapidly sequenced following the first infections discovered in late 2019 \citep{zhu2020novel}. The SARS-CoV-2 genome encodes 29 proteins \citep{gordon2020sars}, of which the spike (S) protein has been most widely studied due to its role in facilitating viral infection \citep{walls2020structure}. The 3-D structure of the S protein has the ability to bind to human ACE2 (angiotensin-converting enzyme 2), thereby gaining entry into cells. Thus, the presence of antibodies that recognize and block the binding region of the S protein can provide immunity against COVID-19, as elicited from prior infection \citep{ju2020human} or vaccination \citep{wang2021mrna}. Infusion of monoclonal antibodies as an early treatment of COVID-19 infection can block disease progression in a similar way \citep{marovich2020monoclonal}. Proteins are often flexible and exhibit dynamic structural changes, such that they may have multiple states that facilitate binding \citep{henzler2007dynamic}. Investigating the S protein's structural flexibility in its key binding regions, especially prior to their interaction with ACE2, can thus reveal insights into the potential efficacy of antibody candidates \citep{dehury2020effect}. Computational methods for doing so would also be valuable more generally in drug discovery applications.

\begin{figure}
	\centering
	\includegraphics[width = \linewidth]{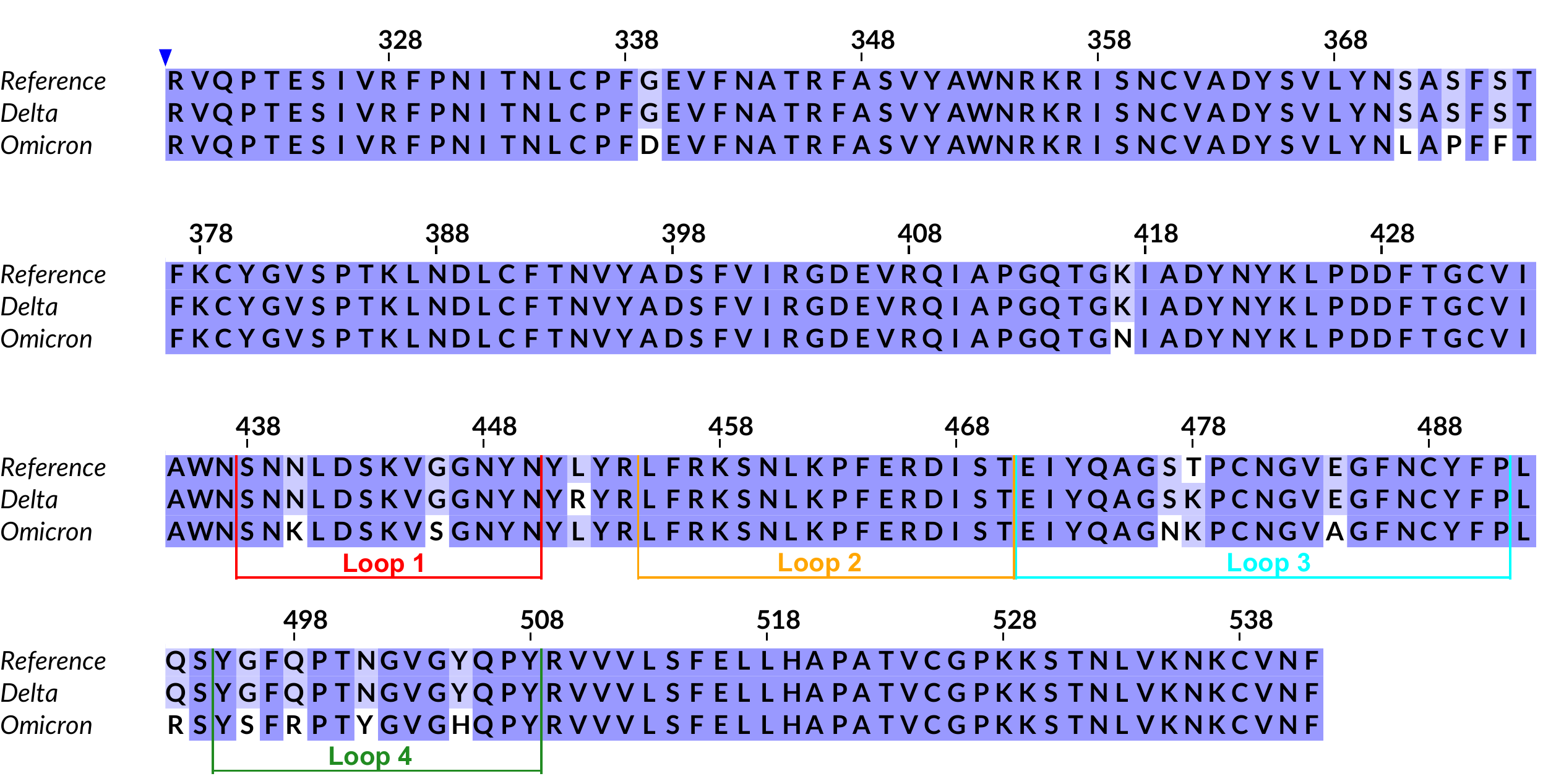} \\
	\includegraphics[width = \linewidth]{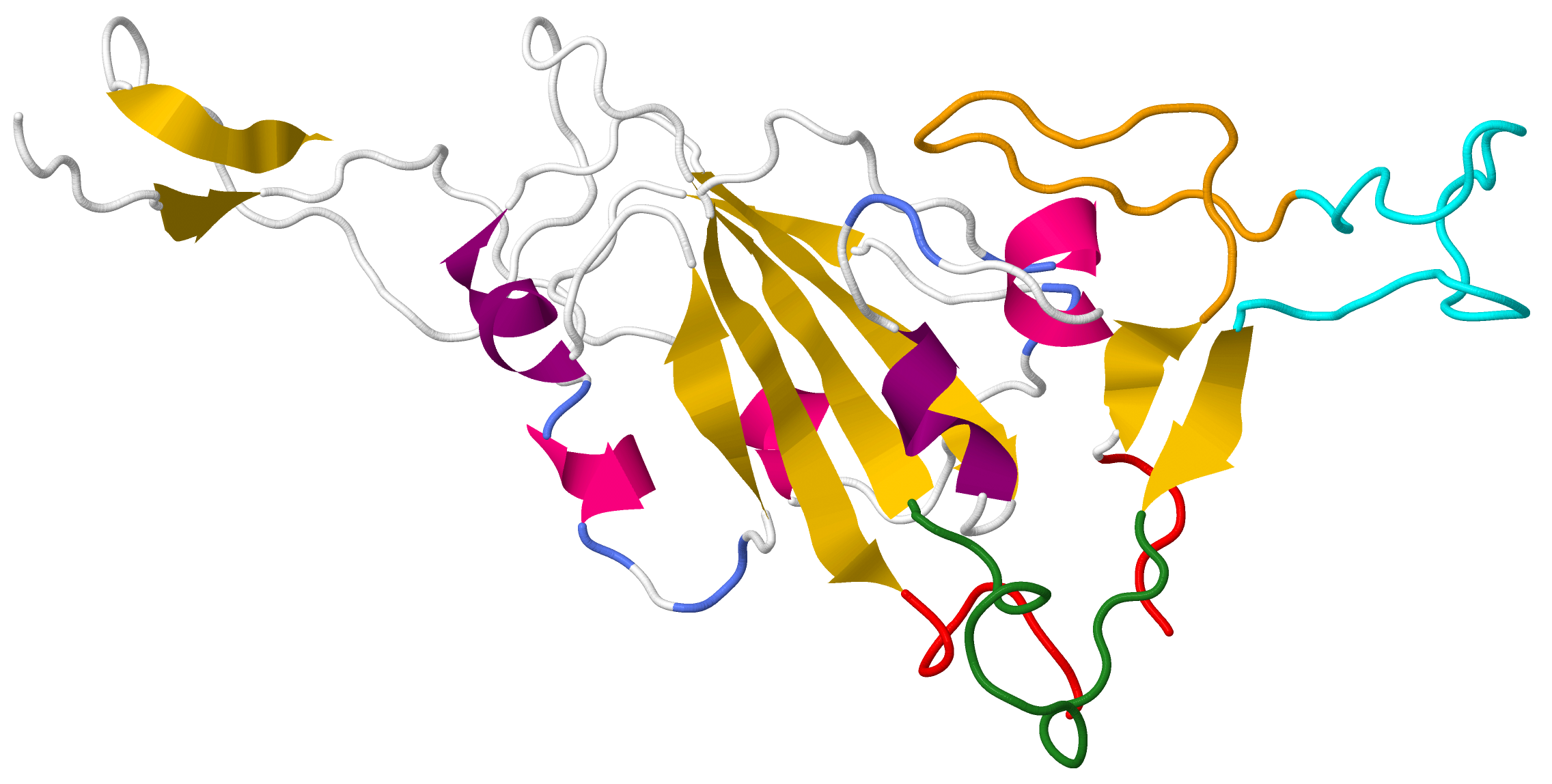}
	\caption{The sequence-to-structure correspondence for the RBD of the SARS-CoV-2 spike protein. The top panel shows the amino acid sequences of the original (reference) protein together with the delta and omicron variants. Light or dark blue shades indicate positions where two or three sequences have the same amino acid type, respectively. The four loops which interact with ACE2 are marked by the colored boxes (red, orange, cyan, and green). The bottom panel shows the laboratory-determined 3-D structure corresponding to the RBD of the reference sequence (PDB code 6XM0). The four loops of interest in the structure are highlighted with the colors corresponding to the top panel. A gap can be seen in the structure within Loop 1 (red), where the laboratory experiment could not resolve the coordinates of the two amino acids in positions 445--446. The RBD also contains several $\alpha$-helices and $\beta$-sheets; these regions have more regular structural pattern compared to loops and are colored with magenta and yellow respectively in the structure.}
	\label{fig_seqstruct}
\end{figure}

The SARS-CoV-2 protein is 1273 amino acids long, of which positions 319--541 are known as the receptor binding domain (RBD) that is responsible for recognizing ACE2 \citep{huang2020structural}. The excerpted RBD from the original (or \textit{reference}) S protein sequence from January 2020 (NCBI identifier: NC\_045512.2) is displayed in the first row of the top panel of Figure \ref{fig_seqstruct}. Within the RBD, there are four key regions with flexible structure (known as \textit{loops}) that have been experimentally observed to directly interact with ACE2 \citep{williams2021molecular}; these are marked by colored boxes and labelled as Loops 1--4 in Figure \ref{fig_seqstruct}. As the pandemic has progressed, mutations in the viral genome have led to variation in the amino acid sequences recorded from new infections; sets of mutations that have been deemed to affect COVID-19 epidemiology are designated as variants of concern by the \citet{whovariant}. These include the delta and omicron variants; to illustrate, their RBD amino acid sequences are aligned with that of the reference sequence in the top panel of Figure \ref{fig_seqstruct}, where mutated positions are indicated by the lighter blue shades. The delta variant, which was predominant for most of late 2021, has only two amino acid mutations in the RBD  (at positions 452 and 478) compared to reference. In contrast, the omicron variant that has become dominant as of December 2021 is characterized by a much larger number of mutations, with 15 in the RBD alone. Some mutations can affect protein structure and function, such that the immunity provided by previous antibodies may no longer be as effective against new variants of SARS-CoV-2 \citep{harvey2021sars}.

As a result of laboratory structure determination efforts, the Protein Data Bank \citep[PDB,][]{bernstein1977protein} has grown to over 185000 entries as of 2021.
Each PDB structure provides a static snapshot of a protein's arrangement of atoms in 3-D coordinates, known as a \textit{conformation}. Since early 2020, laboratories have contributed over 700 PDB entries relating to the SARS-CoV-2 S protein alone. To illustrate the sequence-to-structure correspondence, the bottom panel of Figure \ref{fig_seqstruct} shows a consensus 3-D structure of the reference RBD obtained from the PDB (accession code 6XM0), where the locations of Loops 1-4 in the structure are highlighted using the same colours as the top panel. While this figure only displays a single static snapshot of the protein, analyses of the available PDB structure data show that many loop regions of the S protein have significant conformational flexibility \citep{wong2021conformational}, including Loops 1, 3 and 4. This echos the findings of studies that have used molecular dynamics (MD) simulations to assess the structural flexibility of the RBD, where Loops 3 and 4 were identified to be among the most flexible regions \citep{dehury2020effect};  \citet{williams2021molecular} also identified additional conformations for Loop 3 that were not represented in the PDB. Further, atomic coordinates that cannot be determined by laboratory experiment are missing from published PDB structures, most often occurring in regions with high structural flexibility \citep{shehu2006modeling}; in Figure \ref{fig_seqstruct}, two such amino acids are missing from Loop 1.  Computational approaches can thus complement experimental knowledge and help address some of its limitations; this can be especially useful when new mutations are discovered and PDB structures have not been determined \citep{chen2020mutations}.

The energy landscape theory \cite{onuchic1997theory} has provided a general approach taken by many computational methods for protein structure prediction, namely to find the most probable stable conformation with the minimum potential energy. That approach however is less useful for quantifying the range of structural variation for a loop that is highly flexible and potentially has multiple stable states. For this latter goal, we would ideally draw samples from the Boltzmann distribution $\pi(\bm{x}) \propto \exp[ -H(\bm{x})/T]$, where $\bm{x}$ represents a conformation, $H$ is an energy function, and $T$ is the effective temperature. This type of conformational sampling problem was studied using sequential Monte Carlo by \citet{zhang2007monte} in the context of simplified protein 3-D models. It becomes more challenging to design samplers whose draws follow $\pi(\bm{x})$ for realistic 3-D protein structures. While subsequent studies have extended the sequential chain growth idea to tackle real proteins \citep{tang2014fast,wong2018exploring}, those sampling methods focused on finding low-energy conformations for protein structure prediction, rather than drawing from the Boltzmann distribution. The value of more broadly sampling the conformational space has been recently recognized in the context of flexible loops as a necessary step to reconstruct their energy landscape \citep{barozet2020reinforcement}, although the exhaustive sampling method presented therein did not attempt to compute the likelihood of its samples.

These considerations motivate the two main contributions of this paper. First, we present a sequential Monte Carlo (SMC) method to draw samples from the protein's 3-D conformational space that follow the Boltzmann distribution for a given energy function. This provides a principled statistical approach that is generally applicable for exploring the energy landscape of conformations for flexible protein loops. Second, as an application of timely interest we apply the proposed method on the key loops of the SARS-CoV-2 S protein RBD, using both the reference and omicron sequences as input. This provides insight into how the range of conformations for each loop may change  as a result of the observed mutations, even as the omicron variant continues to be studied experimentally. We summarize the SMC output by applying hierarchical clustering on the samples obtained, and our results indicate that omicron RBD mutations may be most impactful on the possible conformations of Loop 4, while Loops 1 and 3 remain relatively stable before and after mutation.

\section{Method}

We present an SMC method for drawing samples $\bm{x}$ from the Boltzmann distribution $\pi(\bm{x})$ for a given energy function $H$, and a protein loop represented as a continuous segment of amino acids. To achieve this, we design a new sampling scheme that leverages sequential importance resampling \citep{liu2001theoretical} and the protein sample space representation of \citet{wong2018exploring}. We take as given that a template 3-D structure 
can be supplied as input for the method, which governs how the segment is expected to interact with the rest of the protein. This may be obtained from the PDB if available, such as the reference S protein structure partially shown in the bottom panel of Figure \ref{fig_seqstruct}. Then we require the amino acid sequence and positions for the segment of interest, e.g., `SNNLDSKVGGNYN' for Loop 1 of the reference RBD sequence in Figure \ref{fig_seqstruct} (positions 438--450). In general, suppose the segment is $l$ amino acids long, denoted by $a_1, a_2, \ldots, a_l$. Then the goal of SMC is to sample conformations beginning at the C$_\alpha$ atom of $a_1$ and ending at the  C$_\alpha$ atom of $a_l$ in the template structure, so that proper backbone connectivity is maintained with the rest of the template which is held fixed. The method allows for mutations within the segment to be modeled as well, since no existing structure information is needed for the segment being sampled. Computational analyses of mutations can be quite useful in general since it is impractical to apply laboratory structure determination to study all possible mutations \citep{chen2020mutations}.

\subsection{Overview of sampling setup and strategy}

While atom locations are given as 3-D coordinates in the PDB, for sampling it is more convenient to work with the equivalent dihedral angles that represent the relevant geometric degrees of freedom in protein structures. For an individual amino acid, the dihedral angles $(\phi, \psi, \omega)$ and $\bm{\chi}$ determine the backbone and side chain atom coordinates, respectively; $\bm{\chi}$ is a vector of length 0 to 4, depending on the amino acid type. Specifying a complete length $l$ segment requires  $(\phi_i, \psi_i, \omega_i)$, $i=1, \ldots, l-1$ and  $\bm{\chi}_1, \ldots, \bm{\chi}_l$. To satisfy the required backbone connectivity with the template, the actual free backbone dihedral angles for the segment are $(\phi_i, \psi_i, \omega_i)$, $i=1, \ldots, l-3$, with the remaining six backbone dihedrals determined by analytic closure \cite{coutsias2004kinematic}. Our sequential sampling strategy is then to draw $(\phi_i, \psi_i, \omega_i)$ for one amino acid at a time, and evaluate its incremental energy contribution given the previously sampled amino acids. We sample in an order that alternates between adding one amino acid to the left and right ends of the segment: $a_1, a_l, a_2, a_{l-1}, ...$. This is more efficient than sampling in only one direction, since the two ends are more tightly constrained by the rest of the template \citep{wong2018exploring}. Concurrently with 
this backbone sampling,
we also check possible configurations for the side chain atoms of the partial backbone conformation to ensure that they have feasible intermediate placements. After $l-3$ amino acids are constructed, we use the connectivity requirement to place the remaining backbone atoms. A finalized draw for the entire vector of side chain dihedrals ($\bm{\chi}_1, \ldots, \bm{\chi}_l$) is then obtained conditional on the completed backbone.

SMC methods were originally developed for real-time analysis of dynamically evolving systems, so that the target distribution of interest at a given time-step is represented by a properly weighted sample \citep{liu1998sequential}. The framework of SMC is also useful more generally for sampling from complicated high-dimensional distributions, including simulation of biopolymers \citep{liu2001theoretical}, by decomposing into a sequence of propagation and resampling steps \citep{doucet2001introduction}. In its application here, we use these steps to maintain a weighted sample of partial conformations (or \textit{particles}) as each amino acid is added.

For convenience in what follows, let $N$ be the maximum number of particles in the SMC algorithm and denote $y_i = (\phi_i, \psi_i, \omega_i)$, $y_{1:t} = \{ (\phi_i, \psi_i, \omega_i) \}_{i=1}^t$, $\bm{\chi}_{1:t} = (\bm{\chi}_1, \ldots, \bm{\chi}_t )$. Then $\bm{x} \equiv (y_{1:l-3},\bm{\chi}_{1:l})$  represents a complete conformation for the segment, and the sequential sampling approach we described is based on the factorization of the target distribution as 
\begin{eqnarray}\label{eq:factorize}
\pi(\bm{x}) = \pi_1 (y_1) \pi_2 (y_2 | y_1) \cdots \pi_{l-3} (y_{l-3} | y_{1:(l-4)}) \pi_{sc} (\bm{\chi}_{1:l} | y_{1:(l-3)} ),
\end{eqnarray}
where $\pi_1, \ldots, \pi_{l-3}, \pi_{sc}$ are constructed using the incremental energies of partial conformations, as we subsequently describe.

\subsection{Representation of sample space}

The angular sample space for the backbone dihedrals $y_i$ is $(-180^{\circ}, 180^{\circ}]^3$. In practice, $\omega_i$ is generally close to $180^{\circ}$ due to a planar bond and its distribution may be approximated by a normal random variable with mean $180^{\circ}$ and SD $2.75^{\circ}$ based on PDB structures (except for the Proline amino acid type, where the normal mean may be either $0^{\circ}$ or $180^{\circ}$ with probabilities 0.1 and 0.9 respectively), independent of other dihedral angles. Let $p(\omega_i)$ denote this normal (or mixture of normals) density. The angles $\phi_i$ and $\psi_i$ are much more flexible; their joint distribution is known as the Ramachandran plot and has been studied extensively \citep{hooft1997objectively}. To provide sufficient resolution to accurately reproduce the backbones of PDB structures, their values may be discretized using a $5^{\circ}$ by $5^{\circ}$ grid \citep{wong2018exploring}. Thus, we may exhaustively evaluate a maximum of $K$ $(=72 \times 72)$ directions of growth for the pair $(\phi_i, \psi_i)$. Specifically, we use the discretized grid $\{-180^{\circ}, -175^{\circ}, \ldots, 175^{\circ} \}^2$ and let $ \{(\phi^{(k)}, \psi^{(k)})\}_{k=1}^K$ denote the set of unique angle pairs on the grid.

The sample space for the side chain dihedrals $\bm{\chi}_i$ is often given a discrete representation based on the most commonly observed positions in real structures, known as \textit{rotamers}, and lists are tabulated for each of the 18 amino acid types whose side chains have at least one dihedral angle \citep{ponder1987tertiary}. We follow this approach and use the rotamer lists provided in \cite{shapovalov2011smoothed}, which range from three rotamers for Serine's single dihedral angle (with positions $65.6^{\circ}$, $-179.2^{\circ}$, and $-63.8^{\circ}$) to a total of 75 combinations for Arginine's four dihedral angles. Let $R_i$ denote the list of rotamers for the amino acid type corresponding to $\bm{\chi}_i$. In practice we may perturb the rotamers to increase the allowable range of side chain conformations; here we opt to add Normal noise with SD $10^{\circ}$ to the listed rotamer position of the first dihedral angle \citep{wong2018exploring}.

\subsection{Energy evaluation}

We take the energy function $H$ to be given, which has the role of evaluating the relative likelihood of sampled conformations. A typical $H$ will consider features including dihedral angles and atomic interactions, with the latter computed as a function of pairwise distances between atoms \citep{alford2017rosetta}. For sampling, we may assign $H = +\infty$ for conformations with unrealistic angles or distances (e.g., having atoms that clash with each other); within an SMC algorithm such conformations will then be given weights of zero and discarded. In the context of a segment, $H(\bm{x})$ accounts for the conformation of the segment, along with atomic interactions between the segment and the rest of the template. Atomic clashes will occur very frequently if $\bm{x}$ is naively sampled, which further motivates an SMC approach.

To carry out SMC we need to compute incremental energies, so it is convenient to rewrite  (\ref{eq:factorize}) using an energy perspective (without loss of generality, $H$ is scaled so that $T=1$): 
\begin{eqnarray}\label{eq:energysum}
	H(\bm{x}) = H (y_1) + \Delta H (y_2 | y_1) +  \cdots  +  \Delta H (y_{l-3} | y_{1:(l-4)}) + \Delta H (\bm{\chi}_{1:l} | y_{1:(l-3)} ),
\end{eqnarray}
where we use $\Delta H$ to denote incremental contributions, e.g., of angles and atoms corresponding to $y_2$ given an existing placement of $y_1$. More concretely, we take the form $H(y_1) = - \beta_1 \log[  p( \phi_1, \psi_1 ) p(\omega_1)] + \beta_2 f(\bm{b}_1)$, where $p(\phi_1, \psi_1)$ is the probability of the angle grid pair $(\phi_1, \psi_1)$, $\bm{b}_1$ refers to the positions of the four backbone atoms implied by $(\phi_1,\psi_1,\omega_1)$, $f$ computes the energy of their atomic interactions with the template, and $\beta_1,\beta_2$ are coefficients that depend on the particular energy function chosen. Analogously for subsequent backbone contributions we have $\Delta H(y_i | y_{1:(i-1)}) = - \beta_1 \log[  p( \phi_i, \psi_i ) p(\omega_i)] + \beta_2 f(\bm{b}_i | \bm{b}_1, \ldots, \bm{b}_{i-1})$ where $f$ now additionally depends on the interactions between $\bm{b}_i$ and the previously sampled backbone. Note that for $\Delta H(y_{l-3} | y_{1:(l-4)})$, the $f$ term will be $f(\bm{b}_{l-3}, \bm{b}_{l-2}, \bm{b}_{l-1}  | \bm{b}_1, \ldots, \bm{b}_{l-4})$ to incorporate the energy contributions of the remaining backbone atoms for connectivity. Finally, we have the side chain energy $\Delta H (\bm{\chi}_{1:l} | y_{1:(l-3)} ) =  - \beta_3 \log[  p(\bm{\chi}_{1:l})] + \beta_4 f(\bm{s}_{1:l} | \bm{b}_1, \ldots, \bm{b}_{l-1})$, where $p(\bm{\chi}_{1:l})$ is the joint density of the side chain dihedrals, $\bm{s}_{1:l}$ refers to the positions of all the side chain atoms in the segment implied by $\bm{\chi}_{1:l}$, and $\beta_3, \beta_4$ are additional coefficients.

Details of the energy function used for our analyses are provided in the Supplement.

\subsection{Techniques for computational efficiency}\label{sec:sctechnique}

The main computational cost associated with energy evaluation is the calculation of pairwise distances, which is vital for detecting conformations with atomic clashes. Thus, a sensible strategy is to  ``foresee'' which partial conformations will lead to unavoidable clashes, and remove them from consideration. This is most useful when applied to side chains, since they generally contain more atoms than the completed backbone but their energy contribution $\Delta H (\bm{\chi}_{1:l} | y_{1:(l-3)} )$ would otherwise be evaluated only at the very end; completed backbones with no possible side chain placements would result in wasted computational effort.

We thus implement a side chain pre-computation scheme as follows. When sampling $y_1$, we also pre-compute $\Delta H (\bm{\chi}_{1} | \tilde{y}_{1} )$ for all rotamers in the list $R_1$ where $\tilde{y}_{1}$ is the current growth direction being evaluated; we then store the rotamers, up to a maximum of $n_s$ of them, to represent the distribution of $\Delta H (\bm{\chi}_{1} | \tilde{y}_{1} )$ (i.e., if $|R_1| > n_s$, we sample $n_s$ with probabilities $\propto \exp [ - \Delta H (\bm{\chi}_{1} | \tilde{y}_{1} )]$). However, if all rotamers have $\Delta H (\bm{\chi}_{1} | \tilde{y}_{1} ) = +\infty$, we set $H(\tilde{y}_1) = +\infty$ so that $\tilde{y}_1$ will be removed from consideration. In subsequent backbone sampling, 
suppose we have up to $N_s$ joint rotamer positions for $\bm{\chi}_{1:(i-1)}$ representing the distribution of  $\Delta H(\bm{\chi}_{1:(i-1)} | \tilde{y}_{1:(i-1)})$ for a partial backbone $\tilde{y}_{1:(i-1)}$, $i\ge2$. Then when evaluating a growth direction $\tilde{y}_i$ for $y_i$, we similarly pre-compute $\Delta H (\bm{\chi}_{i} |\tilde{y}_{1:(i-1)}, \tilde{y}_{i} )$ for all rotamers in $R_i$ and set $\Delta H (\tilde{y}_{i} | \tilde{y}_{1:(i-1)}) = +\infty$ if all $\Delta H (\bm{\chi}_{i} |\tilde{y}_{1:(i-1)}, \tilde{y}_{i} ) = +\infty$. Otherwise, we store $n_s$ rotamers out of $R_i$, again sampled if necessary $\propto \exp [ - \Delta H (\bm{\chi}_{i} | \tilde{y}_{1:(i-1)}, \tilde{y}_{i}  )]$. Based on the $N_s$ stored positions for $\bm{\chi}_{1:(i-1)}$ and $n_s$ for $\bm{\chi}_{i}$, we can efficiently obtain $N_s n_s$ candidates for $\Delta H (\bm{\chi}_{1:i} |\tilde{y}_{1:(i-1)}, \tilde{y}_{i} ) = \Delta H(\bm{\chi}_{1:(i-1)} | \tilde{y}_{1:(i-1)}) + \Delta H (\bm{\chi}_{i} |\tilde{y}_{1:(i-1)}, \tilde{y}_{i} ) + \Delta H(\bm{\chi}_{1:(i-1)} | \tilde{y}_{i}) + \Delta H (\bm{\chi}_{i} | \bm{\chi}_{1:(i-1)})$, where the first two terms are already pre-computed. Finally, if all these $\Delta H (\bm{\chi}_{1:i} |\tilde{y}_{1:(i-1)}, \tilde{y}_{i} ) = +\infty$ we set $\Delta H (\tilde{y}_{i} | \tilde{y}_{1:(i-1)}) = +\infty$; otherwise we sample $N_s$ from these $\propto \exp [-\Delta H (\bm{\chi}_{1:i} |\tilde{y}_{1:(i-1)}, \tilde{y}_{i} )]$ for use when sampling the backbone of the next amino acid.

The role of $n_s$ and $N_s$ is to achieve a balance between computational cost and maintaining a reasonably accurate representation of $\Delta H (\bm{\chi}_{1:i} |y_{1:i})$ during intermediate steps. This has two practical benefits: (i) partial backbone conformations with no feasible side chain placements are removed early; (ii) the pre-computed joint rotamer positions reduce the time needed to finish side chain sampling at the end (only rotamers for $\bm{\chi}_{(l-2):l}$ will remain to be evaluated). While \cite{wong2018exploring} also proposed keeping a collection of intermediate $\bm{\chi}_{1:i}$, there are two important distinctions: (i) our $\bm{\chi}_{1:i}$ are sampled according to the energy function; (ii) our stored energies of $\bm{\chi}_{1:i}$ are only used to remove dead ends and not for the resampling steps, so that any bias introduced to the weights will be minimal.

\subsection{SMC algorithm} \label{sec:smc}

Based on the preceding setup, the scheme of the SMC algorithm is laid out as follows: 

\begin{enumerate}
	
\item \textit{Initialization.} We initialize the SMC algorithm by constructing the first set of particles representing $\pi_1(y_1)$. For each of the angle pairs of the grid, i.e.,  $(\phi^{(k)}, \psi^{(k)})$, $k=1,\ldots,K$,  we draw an independent value of $\omega^{(k)}$ from $p(\omega_1)$, then set $\tilde{y}_1^{(k)} = (\phi^{(k)}, \psi^{(k)}, \omega^{(k)})$ and compute $\tilde{w}_1^{(k)} = \exp[-H(\tilde{y}_1^{(k)} )] / p(\omega^{(k)})$. Then the feasible directions of growth for the first amino acid are represented by the collection of $(\tilde{y}_1^{(k)},\tilde{w}_1^{(k)})$ where $\tilde{w}_1^{(k)} \ne 0$; suppose there are $M \le K$ of these. We use this collection to define the particles $\{(y_1^{(j)},w_1^{(j)})\}_{j=1}^{M}$ for $\pi_1(y_1)$ after normalizing the weights, i.e., $\sum_{j=1}^{M} w_1^{(j)} = 1$. The set of rotamers and energies for $\Delta H(\bm{\chi}_1 | y_1^{(j)})$ are also stored (Section \ref{sec:sctechnique}). In practice we will choose $N\gg K$, so resampling is not needed and we proceed directly to the propagation step for $\pi_2(y_2|y_1)$.

\item \textit{Propagation of backbone.} Suppose we have a set of $M$ particles  $\{(y_{1:(i-1)}^{(j)},w_{i-1}^{(j)})\}_{j=1}^{M}$ that represent partial backbone conformations, where $2 \le i \le l-3$. Taking $y_{1:(i-1)}^{(j)}$, we add the $i$-th amino acid by appending $\tilde{y}_i^{(jk)} = (\phi^{(k)}, \psi^{(k)}, \omega^{(jk)})$ for each angle pair $(\phi^{(k)}, \psi^{(k)})$ of the grid, $k=1,\ldots,K$ together with an independent draw $\omega^{(jk)}$ from $p(\omega_i)$ and computing  $\tilde{w}_i^{(jk)} = w_{i-1}^{(j)} \exp[- \Delta H(\tilde{y}_i^{(jk)} | y_{1:(i-1)}^{(j)}  )] / p(\omega^{(jk)})$. Suppose there are a total of $M'$ feasible growths with $\tilde{w}_i^{(jk)} \ne 0$, out of all the possible directions starting from existing particles (i.e.,  $M' \le MK $). We use this collection to define an expanded set of intermediate particles $\{( y_{1:i}^{*(t)},w_i^{*(t)}) \}_{t=1}^{M'}$ representing $\pi_i(y_{1:i})$, i.e., where each $y_{1:i}^{*(t)}$ consists of a unique pair  $(y_{1:(i-1)}^{(j)},\tilde{y}_i^{(jk)})$ with $\tilde{w}_i^{(jk)} \ne 0$ and the intermediate weights $w_i^{*(t)}$ are normalized so that $\sum_{t=1}^{M'} w_i^{*(t)} = 1$. The set of joint rotamer positions and energies for $\Delta H (\bm{\chi}_{1:i} |y_{1:i}^{*(t)} )$ are also stored (Section \ref{sec:sctechnique}).

\item \textit{Resampling.} If $M' \le N$ there is no need to resample; we simply set $M=M'$ and assign the intermediate particles and weights to be $\{ y_{1:i}^{(j)},w_i^{(j)} \}_{j=1}^{M}$. Our scheme rapidly increases the number of intermediate particles after each propagation step, so generally a resampling step is needed to reduce the particle population to size $N$. For this purpose we adopt the optimal sampling of \citet{fearnhead2003line} to select $N$ particles for further propagation. Briefly, this proceeds as follows in our context:  (i) solve the equation $N = \sum_{t=1}^{M'} \min(c w_i^{*(t)}, 1)$ to obtain the value $c$; (ii) divide the $M'$ intermediate particles into two groups, where group 1 contains the $N_c$ particles that have $w_i^{*(t)} \ge 1/c$ and group 2 contains the remaining $M'-N_c$; (iii) resample $N-N_c$ particles from group 2 with probabilities proportional to $w_i^{*(t)}$ using stratified resampling; (iv) set $M=N$ and take $\{ (y_{1:i}^{(j)},w_i^{(j)}) \}_{j=1}^{M}$ to be all the group 1 particles (keeping their intermediate weights $w_i^{*(t)}$) and the $N-N_c$ resampled particles from group 2 (with weights reassigned to be $1/c$). The propagation and resampling steps are iterated until we obtain the set of completed backbones for the segment, i.e., after the connectivity constraint is applied to $\{ (y_{1:(l-3)}^{(j)},w_{l-3}^{(j)}) \}_{j=1}^{M}$.

\item \textit{Finalizing the side chains.} After backbone sampling, we have $\{ y_{1:(l-3)}^{(j)},w_{l-3}^{(j)} \}_{j=1}^{M}$ and $N_s$ joint rotamer positions representing the distributions of $\Delta H ( \bm{\chi}_{1:(l-3)} | y_{1:(l-3)}^{(j)})$, for each $j=1,\ldots M$. To complete the side chains for each backbone, we draw one vector $\bm{\chi}_{1:(l-3)}^{(j)} \propto \exp[-\Delta H ( \bm{\chi}_{1:(l-3)} | y_{1:(l-3)}^{(j)})]$, and then successively sample the three remaining side chain rotamers according to $\bm{\chi}_{l-d}^{(j)} \propto \exp[- \Delta H(\bm{\chi}_{l-d} | \bm{\chi}_{1:(l-d-1)}^{(j)},  y_{1:(l-3)}^{(j)}) ]$, $d=2,1,0$, as long as there exists rotamers with $\Delta H(\bm{\chi}_{l-d} | \cdot) < +\infty$. This simple strategy for sampling $\bm{\chi}_{(l-2):l}$ works for most particles to obtain a final conformation without atomic clashes; those that cannot be completed in this way (empirically $\sim$10-25\%) are assigned a final weight of zero.

\end{enumerate}

\section{Application: SARS-CoV-2 omicron variant}

We focus our analyses on Loops 1, 3, and 4, corresponding to the segments 438--450, 471--491, and 495--508 as indicated in Figure \ref{fig_seqstruct}. Compared to the reference sequence, the omicron variant has 2, 3, and 4 mutations in these loops, respectively (no mutations were observed in Loop 2). The full-length S protein from the PDB structure 6XM0 was used as the template. For the main results, we ran the proposed SMC method on the reference and omicron sequences for these three loops, using $N=50000$ particles, $N_s=25$ and $n_s=20$ for side chain pre-computation. We also discuss and compare with existing loop sampling and prediction methods to provide additional context for our results. 

\subsection{Main results}

We use hierarchical clustering as a simple way to summarize the SMC samples, so that the main distinct conformations possible for each loop can be represented by the clusters identified. Complete linkage is chosen to encourage the identification of compact clusters \citep[p. 62,][]{everitt2011cluster}. To compute the distance matrix between sampled conformations, we adopt the most commonly-used metric for comparing protein structures, namely the root-mean-square deviation (RMSD) of the backbone atoms. For each loop, we perform this cluster analysis on the combined SMC samples from the reference and omicron sequences. Then insights into a key question of interest, namely the effects of mutation on the energy landscape of the loop, may be gleaned in terms of the conformational clusters. Specifically, if a cluster is mainly composed of samples from either reference or omicron, this suggests the conformational space represented by that cluster is more favorable for one amino acid sequence than the other. Across the different clusters, we can then see evidence of whether the range of conformations for the loop has changed as a result of mutation.

The results of this cluster analysis for Loops 1, 3, and 4 are displayed in Table \ref{tab:loopresults}, after cutting each dendrogram into 10 clusters. The reference and omicron composition of each cluster is presented in terms of the number of SMC samples (``$n$'' columns) and total particle weight (``Weight'' columns); e.g., cluster 1 for Loop 3 contains 33096 SMC samples, 17546 from reference (with a normalized total weight of 0.425 among reference samples) and 15550 from omicron (with a corresponding total weight of 0.398). For Loops 1 and 3, all 10 clusters have compositions that are distributed relatively evenly between reference and omicron; this suggests that according to the energy function there are no dramatic changes to the accessible conformational space of these two loops as a result of the omicron mutations. In contrast, the results for Loop 4 have more pronounced empirical differences, with the compositions of three clusters in particular being sharply divided between the reference and omicron sequences: omicron is predominant in cluster 1, while reference is predominant in clusters 6 and 10. Both the number and weight of samples in these clusters suggest a similar conclusion: as a result of the omicron mutations, the energy landscape of Loop 4 shifts to favor the conformation represented by cluster 1, and disfavor those represented by clusters 6 and 10.

\begin{table}[ht]
	\centering
	\caption{Results of cluster analysis comparing the reference and omicron conformations sampled by SMC for Loops 1, 3, and 4 of the SARS-CoV-2 S protein. Hierarchical clustering was applied after combining the SMC samples from reference and omicron for each loop. The reference and omicron composition of each resulting cluster is presented in terms of the number of SMC samples (``$n$'' columns) and total particle weight (``Weight'' columns, normalized so that each column of weights sums to 1).}
	\footnotesize
	\begin{tabular}{crrrrrrrrrrrr}
		& \multicolumn{4}{c}{Loop 1} & \multicolumn{4}{c}{Loop 3} & \multicolumn{4}{c}{Loop 4} \\ \cmidrule(lr){2-5}  \cmidrule(lr){6-9} \cmidrule(lr){10-13}
		& \multicolumn{2}{c}{Reference} & \multicolumn{2}{c}{Omicron} & \multicolumn{2}{c}{Reference} & \multicolumn{2}{c}{Omicron} & \multicolumn{2}{c}{Reference} & \multicolumn{2}{c}{Omicron} \\ \cmidrule(lr){2-3}  \cmidrule(lr){4-5} \cmidrule(lr){6-7} \cmidrule(lr){8-9} \cmidrule(lr){10-11} \cmidrule(lr){12-13}
		
		Cluster & $n$ & Weight & $n$ & Weight & $n$ & Weight & $n$ & Weight & $n$ & Weight & $n$ & Weight \\
  \hline
1 & 5818 & 0.151 & 9328 & 0.331 & 17546 & 0.426 & 15550 & 0.398 & 3428 & 0.069 & 13923 & 0.345 \\ 
2 & 4915 & 0.124 & 6453 & 0.187 & 10238 & 0.267 & 14660 & 0.347 & 8651 & 0.271 & 3923 & 0.097 \\ 
3 & 8293 & 0.198 & 8949 & 0.114 & 5241 & 0.158 & 5664 & 0.112 & 3870 & 0.107 & 6006 & 0.144 \\ 
4 & 8242 & 0.192 & 5036 & 0.096 & 2762 & 0.075 & 3283 & 0.070 & 2200 & 0.057 & 4289 & 0.131 \\ 
5 & 5485 & 0.119 & 6074 & 0.106 & 1887 & 0.033 & 1285 & 0.027 & 1812 & 0.068 & 4869 & 0.108 \\ 
6 & 4868 & 0.121 & 2675 & 0.053 & 1426 & 0.029 & 1349 & 0.030 & 6043 & 0.161 & 481 & 0.008 \\ 
7 & 3470 & 0.049 & 3370 & 0.066 & 310 & 0.004 & 299 & 0.004 & 4538 & 0.122 & 1287 & 0.042 \\ 
8 & 1996 & 0.036 & 1456 & 0.037 & 337 & 0.005 & 268 & 0.003 & 939 & 0.022 & 4270 & 0.093 \\ 
9 & 480 & 0.005 & 601 & 0.005 & 190 & 0.002 & 323 & 0.005 & 1839 & 0.063 & 854 & 0.028 \\ 
10 & 526 & 0.006 & 325 & 0.004 & 151 & 0.002 & 164 & 0.002 & 1805 & 0.060 & 344 & 0.003 \\ 		
		\hline
	\end{tabular}\label{tab:loopresults}
\end{table}

\begin{figure}
	\centering
	\includegraphics[width = 0.8\linewidth]{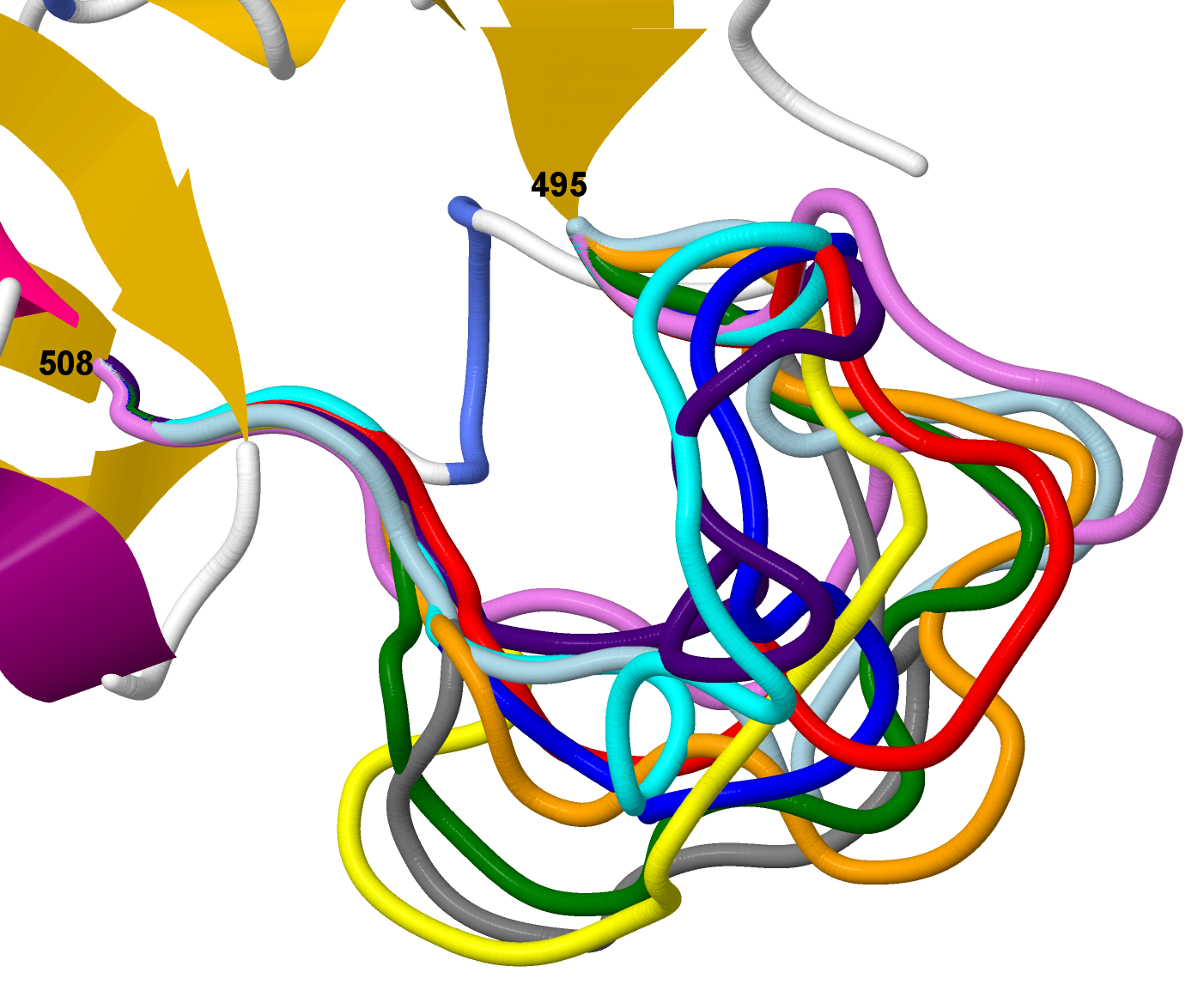}
	\caption{Sampled conformations for Loop 4 based on the reference sequence and the omicron variant. The 3-D structures representing the medoids of the 10 clusters of conformations identified by hierarchical clustering are visualized with different colours. Clusters 1 to 10 corresponding to those listed for Loop 4 in Table \ref{tab:loopresults} are coloured as follows: red, orange, yellow, green, cyan, dark blue, indigo, light violet, grey, and light blue. The starting and ending positions of the loop are labelled (495 and 508). Substantive structural differences can be seen among the clusters. Cluster 1 is predominantly composed of samples from the omicron variant, while clusters 6 and 10 are predominantly composed of samples from the reference sequence.}
	\label{fig_loop4}
\end{figure}

To visualize these conformational clusters, we take the medoid as the representative for each cluster (as the centroid is not guaranteed to satisfy protein geometry). The medoids of the 10 clusters for Loop 4 are plotted in Figure \ref{fig_loop4}, using different colours to illustrate their conformations; e.g., the clusters discussed above are shown in red (cluster 1), dark blue and light blue (clusters 6 and 10). Similar plots for Loops 1 and 3 are provided in the Supplement. With a choice of 10 clusters, the conformations represented are substantively distinct: for Loop 4, the smallest RMSD between any pair of medoids is 1.9 \AA, which is around the 2 \AA~cutoff that has been previously suggested as a threshold for structural similarity \citep{krissinel2007relationship}. The size of the conformational space, and thus the separation between cluster medoids, are generally expected to grow with loop length. Loop 1, which at 13 amino acids long is one shorter than Loop 4, has a somewhat larger 2.3 \AA~between its two closest medoids. An intuitive explanation is that the right end of Loop 4 (near position 508) appears to be more constrained by atomic interactions with the rest of the S protein, such that the 10 clusters have little structural variability there (see Figure \ref{fig_loop4}). Loop 3 is the longest at 21 amino acids; its conformational space is very large and its 10 clusters are all separated by at least 3.8 \AA.

We expect the differences noted between reference and omicron for Loop 4 in Table \ref{tab:loopresults} are larger than can be explained by Monte Carlo variation alone. To informally quantify this, we ran the SMC algorithm twice on the Loop 4 reference sequence and performed the same hierarchical clustering procedure after combining the outputs of the two repetitions. Table \ref{tab:loop4control} shows the cluster compositions by repetition, in a similar format to Table \ref{tab:loopresults}. In relative terms, the number of samples is more stable across the repetitions compared to the total weights. The amount of Monte Carlo variation seen here is comparable to the empirical differences between reference and omicron for Loops 1 and 3, but substantially less than those seen for Loop 4 (clusters 1, 6, 10). A more formal investigation into the variance of SMC in this context would be an interesting direction of future research.

\begin{table}[!htbp]
	\centering
	\caption{Quantifying the Monte Carlo variation across repetitions of the SMC algorithm on the Loop 4 reference sequence. Hierarchical clustering was applied on the combined outputs of two SMC repetitions. The composition of each resulting cluster by repetition is presented in terms of the number of SMC samples (``$n$'' columns) and total particle weight (``Weight'' columns, normalized so that each column of weights sums to 1).}
	\begin{tabular}{crrrr}
		& \multicolumn{2}{c}{Repetition 1} & \multicolumn{2}{c}{Repetition 2} \\  \cmidrule(lr){2-3}  \cmidrule(lr){4-5}
		Cluster & $n$ & Weight & $n$ & Weight \\
		\hline
1 & 8568 & 0.253 & 8970 & 0.215 \\ 
2 & 3656 & 0.116 & 5536 & 0.119 \\ 
3 & 2423 & 0.065 & 2430 & 0.148 \\ 
4 & 3610 & 0.118 & 3719 & 0.086 \\ 
5 & 4065 & 0.127 & 3100 & 0.067 \\ 
6 & 3457 & 0.074 & 3894 & 0.081 \\ 
7 & 2885 & 0.060 & 3334 & 0.088 \\ 
8 & 2459 & 0.058 & 2869 & 0.087 \\ 
9 & 1795 & 0.066 & 1765 & 0.056 \\ 
10 & 2207 & 0.062 & 2149 & 0.053 \\ 		
		\hline
	\end{tabular} \label{tab:loop4control}
\end{table}

\subsection{Comparisons with other methods and approaches}

To provide additional context for our results, we consider some established methods for sampling and prediction of loops: DiSGro \citep{tang2014fast}, next-generation KIC \citep{stein2013improvements}, and SMC-based stochastic optimization \citep{wong2018exploring}. We denote the latter two as NGK and SMCopt for short. Each of these methods is designed to sample the conformational space of a loop and select the lowest-energy conformation found as the prediction. While this approach may be less useful in the context of loops that are flexible enough to adopt multiple stable conformations (see Introduction), we can nonetheless use some simple metrics to assess the samples obtained by these methods in comparison to ours. Software for DiSGro and SMCopt can output a user-specified number of samples in a single run. For NGK (available in the Rosetta macromolecular modelling suite \url{https://rosettacommons.org/}) each run is designed to output a single conformation as a prediction; since each run is stochastic, a larger sample can be obtained by repeated runs of the software. Each of these methods outputs conformations together with their computed energy values (according to the method's energy function), although the samples have not been designed to follow the Boltzmann distribution.

The PDB template contains the ground truth of one known conformation for each of Loops 3 and 4, in the case of the reference sequence.  Samples that resemble this conformation (e.g., as measured by RMSD) should be present among those generated by an effective sampling method. This provides a first metric of comparison: the minimum RMSD to the template conformation among the samples generated by each method. When the goal is to predict a single `correct' conformation for the loop, the usual approach is to choose the lowest-energy sample as the prediction. In the case of highly flexible loops, it would not be surprising if the lowest-energy and template conformations have significant disagreement; this phenomenon has been previously observed in studies of loops with multiple stable conformations \citep{marks2018predicting}. We use this as a second illustrative metric: the RMSD to the template conformation of the lowest-energy sample from each method.

The results for these metrics on Loops 3 and 4 are shown in Table \ref{tab:loopcompare};  to use each method to sample as exhaustively as possible, DiSGro was run to output 100,000 conformations, SMCopt was run with 50000 particles (to match our $N=50000$ particles), and NGK was run 500 times. DiSGro could handle a maximum length of 20 amino acids, so we reduced its Loop 4 task by one amino acid, to 471--490. The minimum RMSD metrics confirm that the samples from our SMC method can cover the conformational space near the template comparably to methods designed for loop prediction.  Overall, we expect the minimum RMSDs to depend on the size of the conformational space that increases with loop length; here the methods obtain $\sim$1-2 \AA~for Loop 3 and $\sim$3-4 \AA~for Loop 4. In contrast, the RMSDs of the lowest-energy sample to the template tend to have much larger values (6--12 \AA, with SMCopt's Loop 3 being an exception), confirming our intuition that these flexible loops may be better studied beyond a traditional loop prediction framework.

\begin{table}[ht]
	\centering
	\caption{Sampled conformations for Loops 3 and 4 using SMC, DiSGro, NGK, and SMCopt. Two metrics are calculated for each method: the minimum RMSD to the template conformation among all its samples; the RMSD to the template conformation of its lowest-energy sample. DiSGro results for Loop 4 are approximated by sampling 471--490 due to the limitations of the method.}
	\begin{tabular}{cccccccccc}
		 & & \multicolumn{4}{c}{Minimum RMSD to template } & \multicolumn{4}{c}{RMSD of lowest-energy sample} \\  \cmidrule(lr){3-6}  \cmidrule(lr){7-10}
		Loop & Positions & SMC & DiSGro & NGK & SMCopt &  SMC & DiSGro & NGK & SMCopt \\
		\hline
		3 & 438--450 & 2.07 & 1.81 & 1.71 & 1.16 & 7.96 & 10.80 & 6.42 & 1.92\\ 
		4 & 471--491 & 3.37 & 3.49& 2.51 & 4.20 & 11.59 & 12.18 & 10.25 & 12.84 \\ 
		\hline
	\end{tabular} \label{tab:loopcompare}
\end{table}

\section{Discussion and conclusion} \label{sec_conclusion}
								   
This paper presented a new application of SMC for sampling the conformational space of loops in protein structures, according to the Boltzmann distribution for a given energy function. Compared to existing methods for loop sampling and prediction that focus on finding the lowest-energy conformation, the emphasis of this work is to broadly explore the energy landscape of possible conformations using principled statistical methodology.  Our approach is especially pertinent for studying loops that can adopt a range of conformations rather than a single stable one. Evidence for such highly flexible loops has been seen in both laboratory experiments that generate PDB structure data, and computational approaches such as MD simulations. As a case study of timely interest, we applied the proposed SMC method to key regions in the RBD of the SARS-CoV-2 S protein that are known to be flexible and have a number of mutations in the omicron variant discovered November 2021.

Our results indicate that the energy landscapes of Loop 4 for the reference and omicron variant cover a different range of conformations, when compared via a cluster analysis of the SMC samples; in contrast Loops 1 and 3 appear to be more stable before and after mutation. Other preliminary studies on the omicron variant have sought to identify the key mutations that may lead to a new or stronger binding interface with ACE2: among the 15 omicron RBD mutations, those identified were 493, 496, 501, 498 \citep{lubin2021structural} and 477, 496, 498, 501 \citep{wang2022structural}. The mutations at 496, 498, and 501 were deemed significant by both studies; all three of these positions are within Loop 4 and agree with our findings on its highest propensity for conformational change. Overall, SMC can be a powerful tool and used in conjunction with other computational analyses of protein structure (e.g., MD simulations, comparative modelling, binding free energy calculations) to obtain new insights.

As some directions for future research, we could investigate more formal ways to use the SMC samples to test for conformational differences between sequence variants, and to quantify the Monte Carlo variance of the SMC method in this context. It would also be of practical interest to apply the method to study more extensive datasets, such as protein loops that are known to be highly flexible \citep{marks2018predicting}, PDB structures with missing loops that cannot be determined by laboratory experiment \citep{shehu2006modeling}, and any new SARS-CoV-2 variants that may continue to emerge.


\subsection*{Acknowledgements}
	
	This work was partially supported by Discovery Grant RGPIN-2019-04771 from the Natural Sciences and Engineering Research Council of Canada.

	
	\bibliographystyle{biom}
	\bibliography{omicron_ref}

    





\clearpage
\appendix
\small
\titleformat*{\section}{\normalsize\bfseries}
\renewcommand{\thetable}{S\arabic{table}}	
\renewcommand{\thefigure}{S\arabic{figure}}	
\setcounter{table}{0}
\setcounter{figure}{0}

\subsection*{Supporting Information for ``Monte Carlo sampling of flexible protein structures: an application to the SARS-CoV-2 omicron variant''}

\noindent This supplement provides additional details on the energy function used for the analysis, plots of the sampled conformational clusters for Loops 1 and 3, and instructions for running the code.

\section{Details of energy function} \label{app:energy}

The energy function used in this paper combines aspects of those previously developed for protein loop sampling and prediction.

For the backbone dihedral density $p(\phi_i, \psi_i)$, we took the empirical probabilities computed by \citet{wong2018exploring} on the same $5^\circ$ by $5^\circ$ grid, for each of the 20 amino acid types based on a database of loop structures. The angle pairs on the grid with very low empirical probability ($<0.00002$) were treated to be zero.  We used a Normal density for $p(\omega_i)$ with mean $180^\circ$ and SD $2.75^\circ$, except for Proline which was taken to be the mixture of Normals $0.9 N(180^\circ, 2.75^\circ) + 0.1 N(0^\circ, 2.75^\circ)$ as described in the  main text.

For the function $f$, the energies of pairwise atomic interactions were taken to be the values derived in the DiSGro study \citep{tang2014fast} for scoring loop conformations, with two modifications. First, we set the energy of pairwise atomic clashes (i.e., when the Van der Waals forces of a pair exceeds 10.0 kcal/mol according to the Lennard-Jones 12-6 model, as in \citet{wong2018exploring}) to be $H = +\infty$. This ensures the final conformation will be free of atomic clashes, compared to the more relaxed maximum pairwise energy of 8.0 used originally in DiSGro. Second, as in \citet{wong2018exploring} we used a distance check between the backbone atoms of the current amino acid $\bm{b}_i$ and the C$_\alpha$ on the other end of the segment and set $H = +\infty$ if it is not possible to complete the segment with realistic bond lengths and angles.

We took the side chain dihedral density $p(\bm{\chi}_{1:l}) \propto 1$, that is, uniform over all allowable rotamer positions. With this choice, the energy contribution of side chain atoms is entirely based on their atomic interactions as computed via $f$.

Finally, values were needed for the coefficients $\beta_1,\beta_2,\beta_3,\beta_4$ that weight the different energy components. Based on the empirical results in \cite{wong2018exploring}, a coefficient ratio of 10 to 1 between the dihedral and DiSGro energy components is a good choice. We then used the loops of length 8 and 12 given in Table 4 of \cite{wong2018exploring} as a tuning set to select the overall energy scaling factor (i.e., equivalent to the effective temperature), taking the minimum RMSD to the template among the SMC samples as our metric. Based on these experiments, we chose $\beta_1 = \beta_3 =  1.0$ and $\beta_2 = \beta_4 = 0.1$ (noting that $\beta_3$ is inconsequential here, since the side chain dihedrals were taken to be uniform).

\section{Plots of conformational clusters for Loops 1 and 3} \label{app:loopplots}

\begin{figure}[!htbp]
	\centering
	\includegraphics[width = 0.9\linewidth]{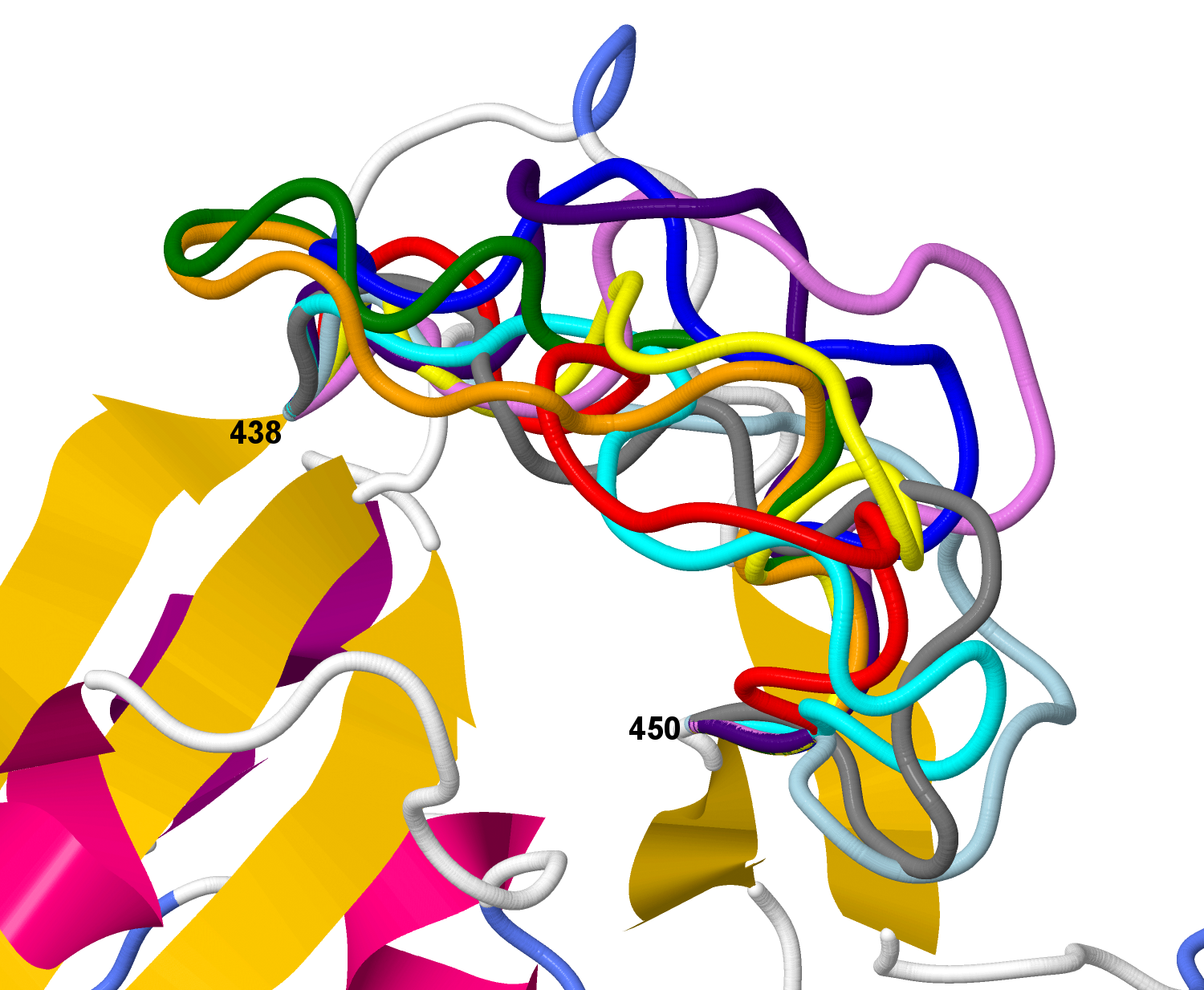}
	\caption{Sampled conformations for Loop 1 based on the reference sequence and the omicron variant. The 3-D structures representing the medoids of the 10 clusters of conformations identified by hierarchical clustering are visualized with different colours. Clusters 1 to 10 corresponding to those listed for Loop 1 in Table 1 of the main text are coloured as follows: red, orange, yellow, green, cyan, dark blue, indigo, light violet, grey, and light blue. The starting and ending positions of the loop are labelled (438 and 450). Clear structural differences can be seen among the clusters. The cluster compositions are distributed relatively evenly between reference and omicron samples.}
	\label{fig_loop1}
\end{figure}

\begin{figure}[!htbp]
	\centering
	\includegraphics[width = 0.8\linewidth]{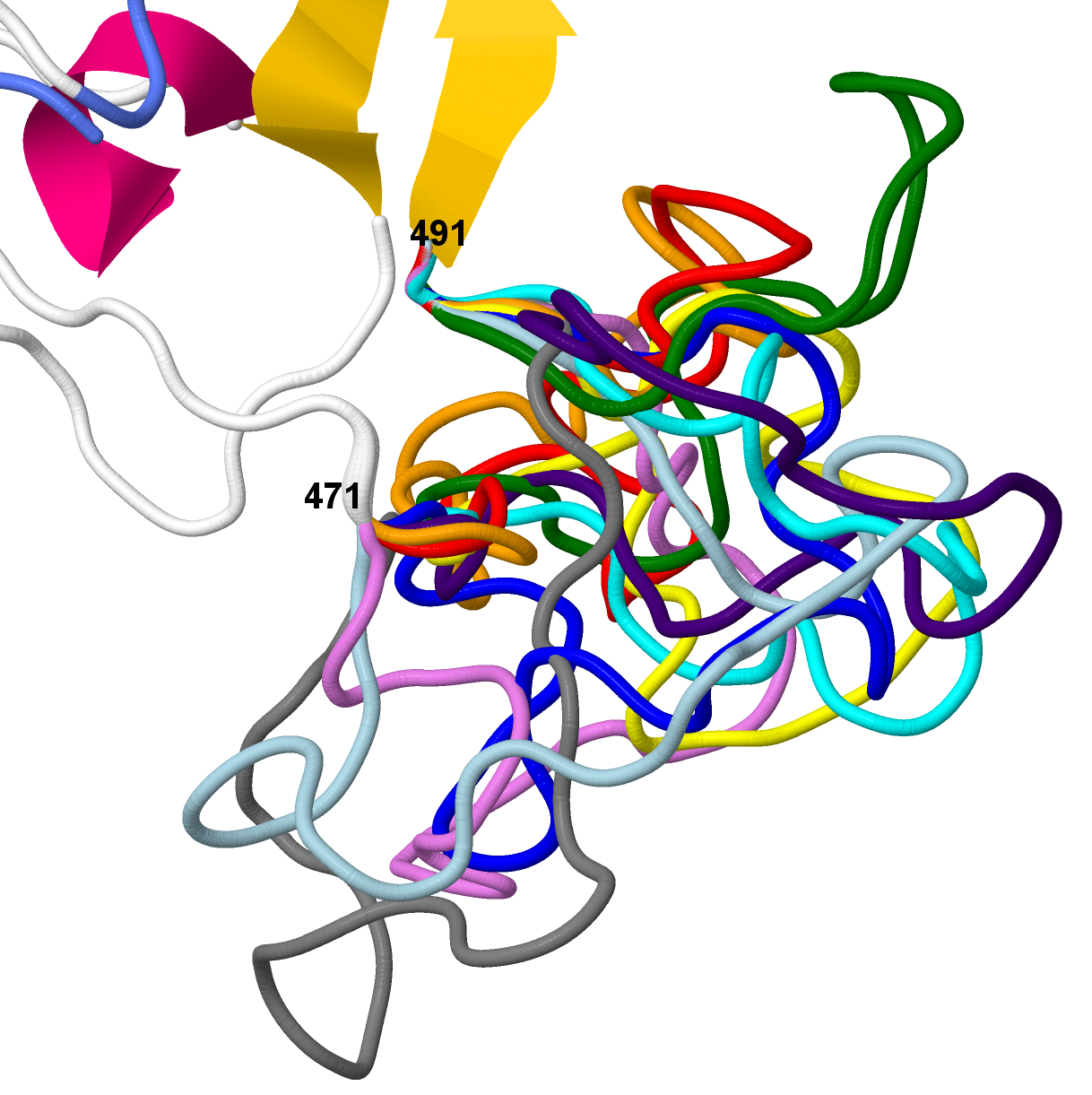}
	\caption{Sampled conformations for Loop 3 based on the reference sequence and the omicron variant. The 3-D structures representing the medoids of the 10 clusters of conformations identified by hierarchical clustering are visualized with different colours. Clusters 1 to 10 corresponding to those listed for Loop 3 in Table 1 of the main text are coloured as follows: red, orange, yellow, green, cyan, dark blue, indigo, light violet, grey, and light blue. The starting and ending positions of the loop are labelled (471 and 491). Clear structural differences can be seen among the clusters. The cluster compositions are distributed relatively evenly between reference and omicron samples.}
	\label{fig_loop3}
\end{figure}

Figure 2 in the main text plots the medoids of the 10 clusters of conformations for Loop 4 of the SARS-CoV-2 spike protein, based on the combined SMC samples from the reference and omicron variant. We display analogous plots for Loops 1 and 3 in Web Figures \ref{fig_loop1} and \ref{fig_loop3}. In contrast to Loop 4, the compositions of the clusters identified for Loops 1 and 3 are more evenly distributed between reference and omicron samples, which suggests the energy landscape is fairly stable before and after mutation. The cluster medoids have substantively distinct structures.  The smallest RMSD between any pair of medoids is 2.3 \AA~for Loop 1; meanwhile Loop 3, which has a much larger conformational space due to its length (21 amino acids), has all its medoids separated by at least 3.8 \AA~RMSD.

\newpage

\section{Running the software} \label{app:software}

A software package for Linux systems that implements the SMC loop sampling algorithm presented in this paper can be downloaded from \url{https://swong.ca/downloads/loopsmc.tar.gz}. The processed data from the Protein Data Bank to supply as reference and omicron structure templates for sampling Loops 1, 3, and 4 are also provided with the package. For detailed usage instructions, see the README file inside the package.

The other loop sampling and prediction methods discussed in Section 3.2 of the main text are also freely available:
\begin{itemize}
	\item DiSGro: package for Linux systems available for download from  \url{http://tanto.bioe.uic.edu/DiSGro/}
	\item NGK: bundled as the program \texttt{loopmodel} in the Rosetta package, for which an academic license can be obtained from \url{https://www.rosettacommons.org}
	\item SMC stochastic optimization: package for Linux systems available for download from  \url{https://swong.ca/downloads/petals-smc.tar.gz}
\end{itemize}

\end{document}